# Pure-Quartic Solitons

Andrea Blanco-Redondo[1*], C. Martijn de Sterke[1], John E. Sipe[2], Thomas F. Krauss[3], Benjamin J. Eggleton[1], and Chad Husko[1]

**Temporal optical solitons have been the subject of intense research due to their intriguing physics and applications in ultrafast optics and supercontinuum generation. Conventional bright optical solitons result from the interaction of anomalous group-velocity dispersion and self-phase modulation. Here we report the discovery of an entirely new class of bright solitons arising purely from the interaction of negative fourth-order dispersion and self-phase modulation, which can occur even for normal group-velocity dispersion. We provide experimental and numerical evidence of shape-preserving propagation and flat temporal phase for the *fundamental pure-quartic soliton* and periodically modulated propagation for the *higher-order pure-quartic solitons*. Using analytic theory, we derive the approximate shape of the *fundamental pure-quartic soliton* exhibiting excellent agreement with our experimental observations. Our discovery, enabled by the unique dispersion of photonic crystal waveguides, could find applications in communications and ultrafast lasers.**

The fascinating phenomenon of optical solitons, solitary optical waves that propagate in a particle-like fashion over long distances [1], has been the subject of intense research during the last decades due to its major role in breakthrough applications such as mode-locking [2], frequency combs [3,4], and supercontinuum generation [5,6] amongst others [7-9]. Temporal solitons in optical media, as studied to date, arise from the balance of the phase shift due to anomalous quadratic group-velocity dispersion (GVD), i.e. $\beta_2=(\partial^2 k/\partial\omega^2)<0$, and the self-phase modulation (SPM) due to the nonlinear Kerr effect [10]. This balance leads to *fundamental solitons* that propagate undistorted. When larger pulse energy is injected, *higher-order solitons* exhibiting recurrent periodic propagation are formed. It has been shown *higher-order solitons* can be used for temporal pulse compression [11].

[1]Centre for Ultrahigh bandwidth Devices for Optical Systems (CUDOS), Institute of Photonics and Optical Science (IPOS), School of Physics, The University of Sydney, NSW 2006, Australia
[2]Department of Physics, University of Toronto, 60 St. George St., Toronto, Ontario, Canada, M5S 1A7
[3]Department of Physics, University of York, York, YO10 5DD, UK.
*E-mail: ablanco@physics.usyd.edu.au

In practice, higher-order nonlinear and dispersive effects often perturb this behavior. In silicon (semiconductor) waveguides the most significant higher-order nonlinearities are associated with free carriers (FCs) generated by two-photon absorption (TPA) [12,13], which have hampered the observation of soliton-based effects in this material. Recently, some of us achieved *higher-order soliton* compression of picosecond pulses in silicon [14] by using a dispersion-engineered photonic crystal waveguide (PhC-wg) with a strong negative $\beta_2$, slow-light-enhanced Kerr nonlinearity, and the careful selection of the pulse duration. Turning to higher-order dispersive effects, the presence of third order dispersion (TOD — $\beta_3=\partial^3 k/\partial\omega^3$) leads to soliton instability [15], whereas positive fourth-order dispersion (FOD — $\beta_4=(\partial^4 k/\partial\omega^4)>0$), can give rise to radiation at specific frequencies [16]. In the presence of negative FOD ($\beta_4<0$), solitons can be stable [17-19]. These studies [17-19] have led to the concept of *quartic solitons*, solitary pulses resulting from the interaction of anomalous GVD and SPM but distorted by the presence of FOD. *Quartic solitons* play a role in important applications such as supercontinuum generation in the mid-infrared [20], ultra-fast lasers [21], and generally any nonlinear device operated close to the frequency of maximum/minimum GVD, where the TOD vanishes.

In this paper we report the discovery of an entirely new class of solitons originating exclusively from the interaction of negative FOD and SPM, which can occur even when the GVD vanishes or is normal. Since such solitons arise just from quartic dispersion and SPM, and to distinguish them from the solitary waves studied in [17-20], we propose the name of *pure-quartic soliton* for this new class of solitary wave. The basic mechanism of *pure-quartic solitons* described above is conceptually depicted in Fig. 1 (a).

This experimental discovery of *pure-quartic solitons* was enabled by the unique dispersion properties of PhC-wgs, which combine very large negative $\beta_4$ with small positive $\beta_2$ for the wavelength under study. We performed time- and phase-resolved propagation measurements using a frequency-resolved electrical gating (FREG) apparatus, which can be modeled using a generalized nonlinear Schrödinger equation (GNLSE). We show shape-preserving propagation and flat temporal phase for *fundamental pure-quartic solitons* and temporal compression and convex nonlinear phase for *higher-order pure-quartic solitons*. Finally, we derive the approximate shape of *fundamental pure-quartic solitons* using analytic theory.

## Results

**Experimental signatures of *pure-quartic solitons***

For demonstrating the existence of *pure-quartic solitons* we used the 396-µm-long dispersion-engineered slow-light silicon PhC-wg [22] shown in the inset of Fig. 1 (b) (see the Methods section). Figure 1(b) shows the waveguide dispersion measured using an interferometric technique [23]. At the pulse central wavelength, 1550 nm, the PhC-wg has a measured group index of $n_g$=30, a GVD of $\beta_2$=+1 ps$^2$/mm, corresponding to *normal* dispersion, a TOD of $\beta_3$=+0.02 ps$^3$/mm, and a FOD of $\beta_4$=-2.2 ps$^4$/mm.

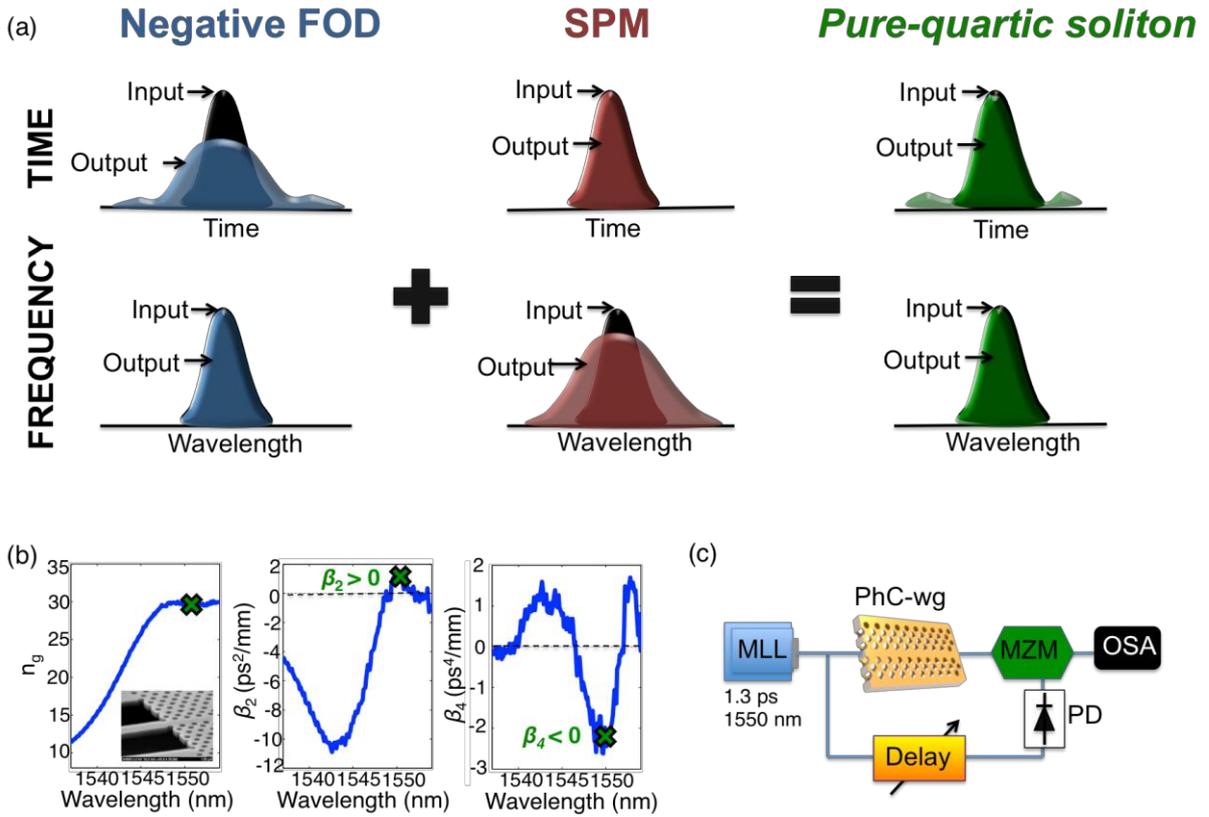

Fig. 1. Concept of *pure-quartic solitons* and their experimental demonstration. (a) Schematics of *pure-quartic solitons*: (Left) Fourth-order dispersion (FOD) gives rise to temporal pulse broadening (blue output pulse vs. black input pulse in time) without affecting the spectrum; (Centre) self-phase modulation (SPM) generates spectral broadening (red output pulse vs. black input pulse in time) without affecting the temporal pulse shape; (Right) the interplay of FOD and SPM can give rise to *pure-quartic solitons* which remain nearly unperturbed (green output pulses vs. black input pulses in both frequency and time). (b) Measured dispersion of the silicon photonic crystal waveguide used in our experiments: group index ($n_g$), second-order dispersion parameter ($\beta_2$) and fourth-order dispersion parameter ($\beta_4$); The inset is a scanning electron micrograph of the sample. (c) Frequency-resolved electrical gating setup: mode locked laser (MLL), photonic crystal waveguide (PhC-wg), tunable delay, ultra-fast photodiode (PD), Mach-Zehnder modulator (MZM), and optical spectrum analyzer (OSA).

In order to perform a complete temporal and spectral characterization of the sub-picojoule ultra-fast nonlinear dynamics in the waveguide we used a FREG apparatus [24] in a cross-correlation configuration, as schematically depicted in Fig.1 (c) (see the Methods section). This setup provides a series of spectrograms, i.e. the gated optical power versus delay, for varying input powers. From these spectrograms we then extract the optical pulses' electric field envelope and phase using a numerical algorithm [25]. Figure 2 shows the measured intensity (red dashed lines) and phase (black dashed) at the output of the PhC-wg, when injecting 1.3 ps Gaussian pulses (full-width at half-maximum, FWHM) at 1550 nm with different input peak powers, $P_0$. Figure 2 (a) shows the frequency domain and Fig. 2 (b) shows the temporal domain. The physical length scales of the dispersion orders for this pulse duration are: $L_{GVD}=T_0^2/|\beta_2|=0.615$ mm, $L_{TOD}=T_0^3/|\beta_3|=22.6$ mm, $L_{FOD}=T_0^4/|\beta_4|=0.168$ mm, with $T_0$=FWHM/1.665 for Gaussian pulses. These length scales indicate that FOD is dominant, with the total length of the sample being $L=2.4 \cdot L_{FOD}$ and the GVD length being $L_{GVD}=3.66 \cdot L_{FOD}$. TOD is negligible in this sample for our pulses.

To understand the origin of the experimental observations in Fig. 2 we employ a GNLSE model to describe the propagation in the silicon PhC-wg:

$$\frac{\partial A}{\partial z} = -\frac{\alpha_{l,eff}}{2}A - i\frac{\beta_2}{2}\frac{\partial^2 A}{\partial t^2} + \frac{\beta_3}{6}\frac{\partial^3 A}{\partial t^3} + i\frac{\beta_4}{24}\frac{\partial^4 A}{\partial t^4} + (i\gamma_{eff} - \frac{\alpha_{TPA,eff}}{2})|A|^2 A + \left(ik_0 n_{FC,eff} - \frac{\sigma_{eff}}{2}\right)N_c A. \quad (1)$$

Here $A(z,t)$ is the slowly varying amplitude of the pulse, $\alpha_{l,eff}$ denotes the linear loss, $\gamma_{eff}$ and $\alpha_{TPA,eff}$ are the effective nonlinear Kerr and TPA parameters, respectively; $n_{FC,eff}$ and $\sigma_{eff}$ represent the free-carrier dispersion (FCD) and the free-carrier absorption (FCA) effective parameters for a free-carrier concentration of $N_c$. Since we use a slow-light PhC-wg, the effective coefficients vary with the slow-down factor $S=n_g/n_0$ [13]. We use the measured envelope amplitude of the input pulse as the input to our GNLSE model with the sample parameters detailed in the Methods section. As shown in Fig. 2, the model (solid lines) agrees well with the experimental data (dashed lines) in both frequency and time.

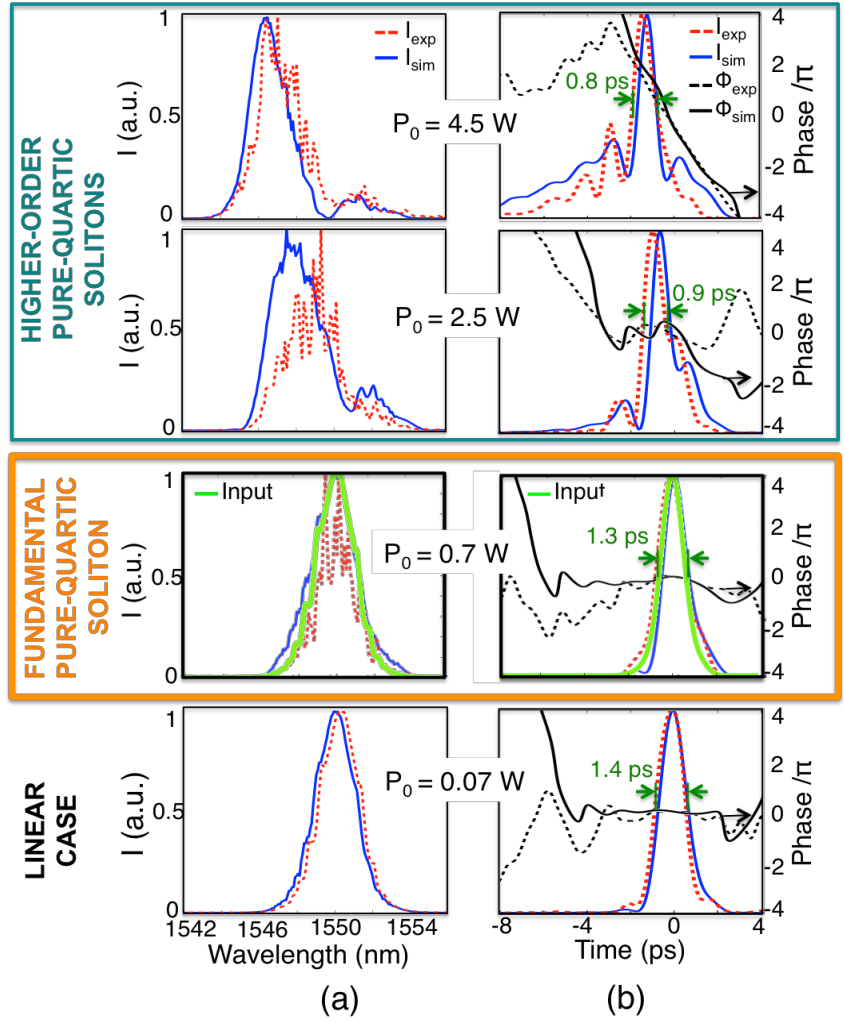

Fig. 2. Experimental and modeling results. (a) Frequency and (b) time domain results for different input powers. The dashed red lines represent the intensity measurements, the blue solid lines represent the intensity simulations, the black dashed line represents the measured phase, and the solid black line represents the simulated phase. The green solid line at 0.7 W represents the normalized input intensity. The yellow box encompasses the *fundamental pure-quartic soliton*, showing nearly unperturbed propagation and flat temporal phase. The turquoise box includes two cases of *higher-order pure-quartic solitons*, showing temporal compression and nonlinear spectral broadening. The *higher-order pure-quartic solitons* observed here are greatly perturbed by the presence of free carriers.

We first focus on the time domain results of Fig. 2 (b). At the low coupled power of 0.07 W, nonlinear effects can be neglected and we simply observe small temporal broadening, mainly due to quartic dispersion. The different signs of $\beta_2$ and $\beta_4$ counteract each other to some degree, leading to a modest temporal broadening at the output of this short PhC-wg (from 1.3 ps to 1.4 ps). We have verified by running the GNLSE for longer lengths that the pulse width keeps increasing with the propagation distance in the linear case. Increasing the input power up to 0.7 W, where the nonlinear length $L_{NL}=1/(\gamma_{eff}P_0)$ becomes comparable to $L_{FOD}$, the pulse preserves its initial shape and duration, as illustrated by the good matching between the measured output intensity (dashed red line) and the normalized input intensity (green solid line). Furthermore, the temporal phase across the pulse duration is nearly flat. These are two signatures of *fundamental soliton* behavior [10]. At 2.5 W, the phase becomes convex due to the stronger nonlinear Kerr effect and the main peak of the pulse narrows, temporal signatures of a *higher-order soliton*. At even higher powers, 4.5 W, the main peak of the pulse narrows even more, corresponding to a *higher-order soliton* with a higher soliton number [11]. In addition, a long tail develops towards the leading edge of the pulse. We provide an explanation for this effect below.

Next we examine the frequency domain in Fig. 2 (a). At 0.07 W, since the nonlinearities are negligible and the pulse spectrum is not affected by the dispersion, the pulse spectral shape is maintained. At 0.7 W the pulse preserves its initial spectral shape, even if the nonlinear effects are significant, again consistent with *fundamental soliton* behavior in the spectral domain. At higher powers (2.5 W and 4.5 W) the pulse experiences spectral broadening and splits into two peaks, spectral signatures of *higher-order solitons*. The observed blue shift and asymmetry are associated mainly to FCD.

Whereas we previously reported shape-preserving *fundamental solitons* and *higher-order soliton* compression in silicon [14], such behavior was unforeseen for the normal GVD here. By setting $\beta_2$=0 in our numerical model we find that the signatures of soliton behavior are maintained: the shape is preserved and the phase is flat for the *fundamental soliton* at 0.7 W, and at 2.5 W and 4.5 W, *higher-order soliton* undergo nonlinear temporal narrowing. This demonstrates that GVD is not important in this system and, since we established that TOD is also negligible, that the soliton behavior stems purely from the interaction of FOD and SPM. Furthermore, we have verified that the long tail at the leading edge observed at high powers (see Fig. 2. (b)), as well as the self-acceleration of the pulse, originate from the interaction of negative FOD and FCD. The FCD generates additional blue-components (see Fig. 2 (a)) and the negative FOD makes them travel faster than the red-components of the pulse, analogous to our earlier results with negative GVD [13,14].

These observations at the output of the silicon PhC-wg suggest the existence of a new type of soliton: *pure-quartic solitons*. We use the term *soliton* here to refer to solitary optical waves that propagate essentially unperturbed over long distances, not to exact localized solutions of integrable nonlinear differential equations [26].

As expected in a silicon system at 1550 nm, *pure-quartic solitons* are strongly perturbed by TPA and FC as we just described, and thus the measured behavior differs from the simple case with just SPM and FOD. Therefore, to elucidate the dynamics of *pure-quartic solitons* in the absence of higher-order nonlinearities we next numerically study the propagation of picosecond pulses along the PhC-wg neglecting all effects but SPM and FOD.

**Propagation behavior of *pure-quartic solitons***

Figure 3(a) and (b) depict the propagation dynamics of undistorted *pure-quartic solitons*, i.e., in the presence of SPM and FOD only. Such a system is governed by the biharmonic nonlinear Schrödinger equation

$$\frac{\partial A}{\partial z} = i\frac{\beta_4}{24}\frac{\partial^4 A}{\partial t^4} + i\gamma_{eff}|A|^2 A. \qquad (2)$$

In our simulations we consider two different power levels: *fundamental pure-quartic solitons* occur at moderate powers (Fig. 3 (a)), whereas at high powers *higher-order pure-quartic solitons* result (Fig. 3 (b)).

The simulations in Fig. 3 (a) show shape-preserving pulse propagation in time and frequency over five quartic dispersion lengths $L_{FOD}$ for a *fundamental pure-quartic soliton*. The output, represented by the blue curve to the right of the propagation plot, shows that the pulse maintains essentially the same amplitude, shape (Gaussian), and duration (1.3 ps) as the input pulse. Importantly, the temporal phase at the output, represented by the black solid line, is flat across the duration of the pulse.

Figure. 3 (b) reveals a *higher-order pure-quartic soliton*, with the pulse experiencing periodic recurrent propagation. In time the pulse undergoes compression and then periodically returns to its initial shape. In frequency the pulse splits into two and then recombines to recover its initial spectral shape after the same period. At the maximum compression point, the pulse reaches the minimum duration of 0.5 ps, a compression factor of 2.6 compared to the initial pulse duration, with a peak intensity of 2.1 times that of the initial pulse. Our simulations of the compression factor and soliton period of the *pure-quartic soliton* follow the same trends as for conventional solitons [11]. Specifically, larger intensities lead to higher compression factors and to compression occurring at an earlier spatial position along the waveguide.

To understand why the experimental observations of *pure-quartic solitons* in Fig. 2 differ from the numerical results of the undistorted system in Fig. 3 (a) and Fig. 3 (b), we simulate the propagation along the PhC-wg including all the effects in the real system indicated in Eq. 1. The outputs shown in Fig. 3 (c) and Fig. 3 (d) match our experimental measurements at $P_0$=0.7W and $P_0$=6.14W. Figure 3 (c) shows that the signatures of the *fundamental pure-quartic soliton* remain in realistic simulations: the pulse maintains its shape and width, and the phase at the output remains almost flat. However, the intensity of the pulse decreases due, predominantly, to the linear loss in the slow-light waveguide ~70 dB/cm). Since the intensity decreases as the pulse propagates, the linear FOD will eventually dominate over the SPM for longer distances, leading to temporal broadening of the *fundamental pure-quartic soliton*. The *higher-order pure-quartic soliton* in the realistic scenario of Fig. 3 (d) differs considerably from Fig. 3 (b). The TPA clamps the intensity in the waveguide from the early stages of propagation. The FCD introduces blue components that lead to the self-acceleration of the pulse and an asymmetry, and the FCA induces absorption on the trailing edge. These effects of TPA and FCs on the propagation of *higher-order pure-quartic solitons* in silicon are analogous to the effects of FCs on conventional solitons [14].

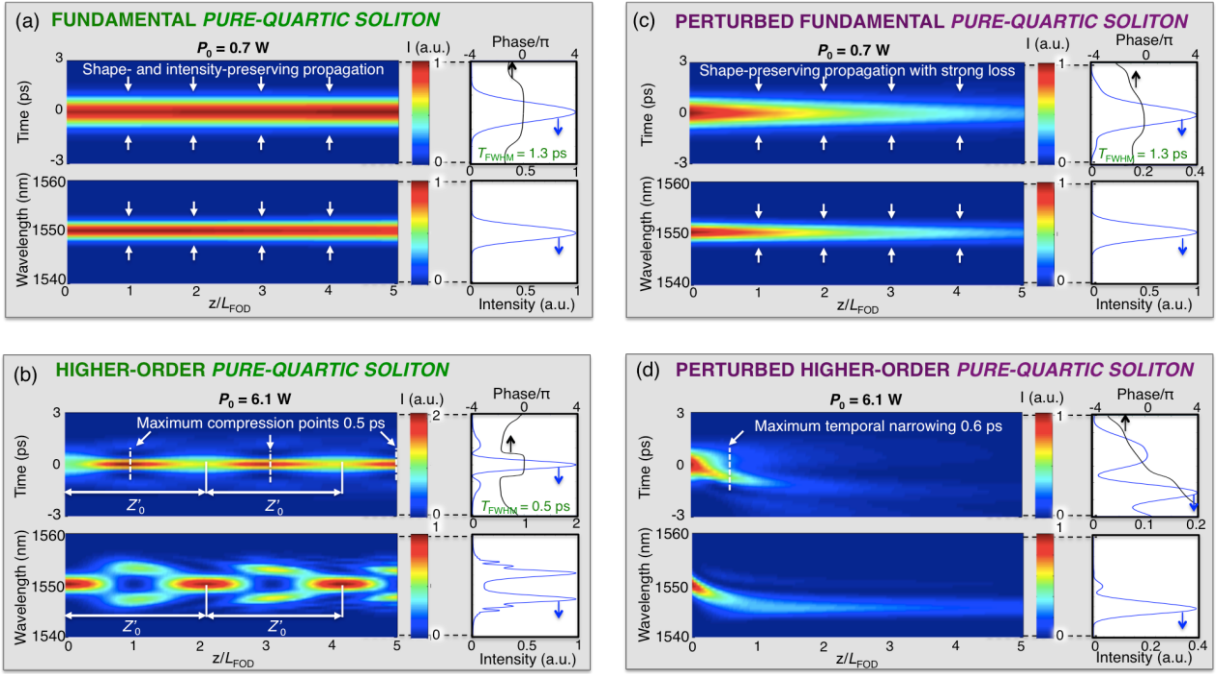

Fig. 3. Simulations of the propagation of a *fundamental* and a *higher-order pure-quartic soliton* along five quartic dispersion lengths, $L_{FOD}$. (a) *Fundamental* ($P_0$ = 0.7W) *pure-quartic soliton* with only self-phase modulation and quartic dispersion present, and (c) in the more realistic scenario for our silicon waveguide with two-photon absorption and free carriers; (b) and (d) are similar but for a higher power level ($P_0$ = 6.14 W) where a *higher-order pure-quartic* soliton results.

**Approximate solution for the *fundamental pure-quartic soliton***

We now discuss an analytic expression for the *fundamental pure-quartic soliton*. The experimental observations and numerical simulations indicate that the central part of *fundamental pure-quartic solitons* appears to be Gaussian. Assuming this shape, we look for an approximate solution to Eq. (2) in two separate ways for verification purposes: using the variational principle and looking for a local approximation near the center. These derivations are described in the Supplementary Material.

In both cases, after a simple dimensional analysis, we take the central part of the *pure-quartic soliton* to be of the form

$$A(z,t) = A_0 e^{i\mu\gamma_{\text{eff}}A_0^2 z} e^{-\left(A_0\sqrt{\nu\sqrt{\frac{\gamma_{\text{eff}}}{|\beta_4|}}}\right)t^2}, \tag{3}$$

where, since $\beta_4 < 0$, we have written $\beta_4 = -|\beta_4|$ for convenience, and $\mu$ and $\nu$ are free parameters. The variational approach gives $\mu = 7/(8\sqrt{2}) \approx 0.62$, $\nu = 1/\sqrt{2}$, whereas the local approximation gives $\mu = \frac{1}{2}$, $\nu = 1$. Thus these approaches predict pulse widths which differ only by a factor $2^{\frac{1}{8}}$ or by less than 10%. The fact that these different approximations give very similar results reinforces our confidence in them. Based on these results, the argument of Akhmediev and Karlsson [27] suggests that *pure-quartic solitons* do not lose energy due to linear radiation since the linear and nonlinear dispersion relations do not intersect. We note Eq. (3) can be written in terms of the normalized length scales we introduced earlier:

$$A(z,t) = A_0 e^{i\mu\frac{z}{L_{\text{NL}}}} e^{-\left(\sqrt{\nu\frac{L_{\text{FOD}}}{L_{\text{NL}}}}\right)\left(\frac{t}{T_o}\right)^2}. \tag{4}$$

To test the validity of this analytic approximate solution we numerically solve Eq. (2) with the $\beta_4$ and $\gamma_{\text{eff}}$ of our sample (see Methods) and obtain the output of the system at the power level corresponding to a *fundamental pure-quartic soliton* for a propagation length $L=30 \cdot L_{\text{FOD}}$. This long propagation distance ensures convergence of the pulse evolution. The results of this numerical experiment for three different pulse shapes: a Gaussian, a hyperbolic secant (sech), and a Super-Gaussian of order 4, are depicted by a solid blue curve in Fig. 4 (a), Fig. 4 (b), and Fig. 4 (c), respectively. Importantly, the hyperbolic secant and super-Gaussian inputs (black solid lines in Fig. 4 (b) and Fig.4 (c) respectively) evolve into the solitary-wave Gaussian shape, constituting an additional signature of soliton-like behavior and proving that this type of soliton acts as an attractor. These results are overlapped with the variational approximation (dashed red line) and the local approximation (green dashed line), using the same parameters. The agreement between the numerical and the variational solution in the central part of the pulse is remarkable. The local approximation deviates only slightly. In addition we numerically find that $\mu \approx 0.63$, again very close to the variational result. The wings observed in the numerical solution relate to the fact that the phase shift profile of the FOD has a quartic dependence with time, whereas the SPM varies quadratically, as illustrated in Fig. 4 (d). This allows both phase shift profiles to perfectly counterbalance each other close to the center of the pulse, but deviate from each other at the edges. This fact is not captured in the approximate analytic solution since a perfect balance between FOD and SPM is assumed.

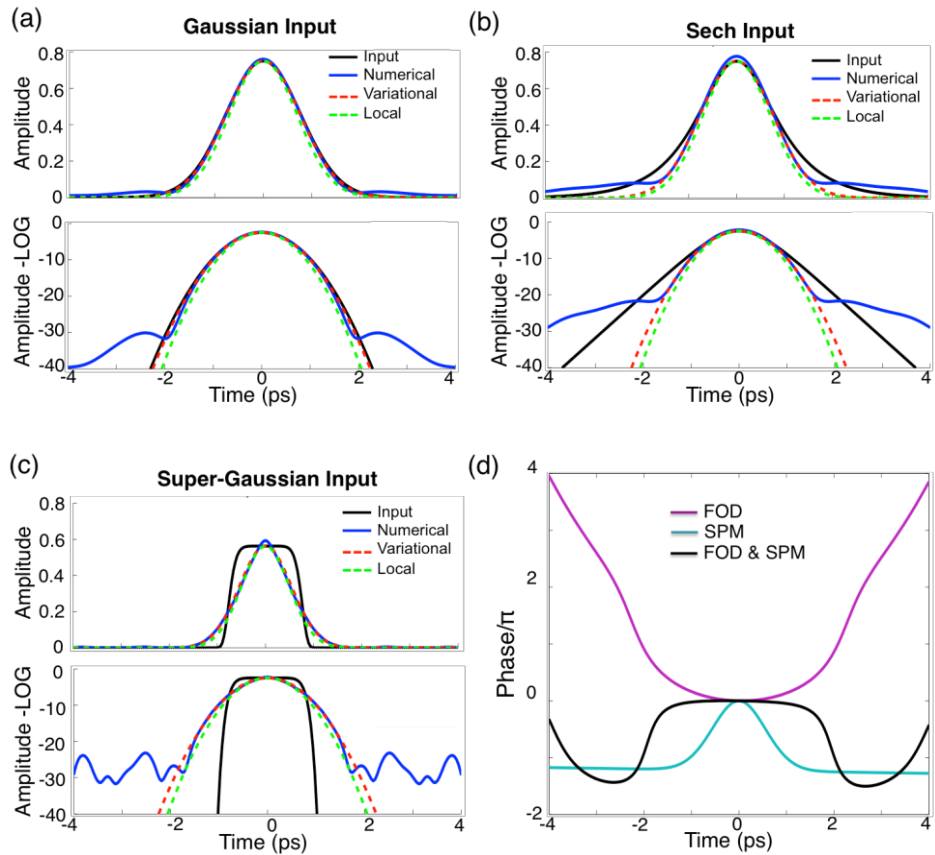

Fig. 4. Approximate solutions to a *fundamental pure-quartic soliton* and phase diagram. Comparison between the variational and local approximate solutions for the *fundamental pure-quartic soliton* and the numerical output after propagating over thirty quartic dispersion lengths for (a) a Gaussian-input, (b) a hyperbolic secant input, and (c) a super-Gaussian input of order 4. (d) Phase shift induced by the fourth-order dispersion (purple) and the self-phase modulation (turquoise) independently and its combined phase shift (black).

## Discussion

The experimental results and the numerical simulations presented here have established the existence of a new class of solitons: *pure-quartic solitons*, arising from the interaction of SPM and FOD only. In particular, we experimentally demonstrated shape-preservation and flat-phase behavior for the *fundamental pure-quartic soliton*, and temporal compression for the *higher-order pure-quartic solitons*. We numerically demonstrated that the *higher-order pure-quartic soliton* would undergo recurrent periodic propagation in the absence of loss and higher-order nonlinearities. Although we have verified that these signatures of soliton propagation are preserved for long propagation distances in the presence of just FOD and SPM, the disparity between the quartic profile of the FOD-induced phase shift and the quadratic profile of the SPM-induced phase shift affecting the edges of the pulse could lead to stability issues that should be studied.

Our discovery was made possible by the unique dispersion properties of PhC-wgs that combine extremely large negative values of $\beta_4$ with normal GVD. However, analogous regimes of propagation could be achieved in highly nonlinear fibers (HNLFs) [28], photonic crystal fibers [8] or specially designed silicon waveguides [20,29]. Preliminary calculations show that, for instance, HNLFs could have wavelength regions in which $\beta_2>0$, $\beta_4<0$, and $L_{FOD} \ll L_D$, $L_{TOD}$, for short pulses ($T_0 \approx 50$ fs). The propagation of such pulses at reasonable peak powers ($P_0 \approx 100$ W) in HNLFs would lead to the formation of *pure-quartic solitons*.

The variational solution here constitutes a good approximation to the central form of the *fundamental pure-quartic soliton*. This approximate solution, valid only for $\beta_4<0$, could stimulate new efforts in finding solutions to the biharmonic nonlinear Schrödinger equation [30,31] also of interest in the field of spatial solitons [32,33]. Furthermore, it would be interesting to investigate analytic solutions supported by the *pure-quartic soliton* system including the effects of linear loss, TPA and FCs.

Aside from their different physical origin, *pure-quartic solitons* present significant differences with respect to conventional solitons. As mentioned, *pure-quartic solitons* open the door to soliton functionality in the normal GVD regime of optical media. More importantly, perhaps, the energy of conventional solitons scales like $(T_0)^{-1}$, whereas the energy of *pure-quartic solitons* scales like $(T_0)^{-3}$, which suggests that they are more energetic for ultra-short pulses. This could entail a potential advantage in the development of soliton lasers, since soliton lasers generating ultrafast pulses are currently limited by a compromise between pulse duration and energy. Therefore, *pure-quartic solitons* are not only of fundamental significance but could also present important applications in communications, metrology and ultrafast lasers.

## Methods

**Device and linear characterization.** The present experiment was performed using a silicon photonic crystal air-suspended structure with a hexagonal lattice (p6m symmetry group) constant $a$ = 404 nm, a hole radius $r$ = 116 nm, and a thickness $t$ = 220 nm. A 396 μm-long dispersion-engineered PhC-wg was created by removing a row of holes and shifting the two innermost adjacent rows 50 nm away from the line defect. The air-clad devices were fabricated with a combination of electron beam lithography, reactive ion and chemical wet etching. The measured linear propagation loss in this slow light region is ~70 dB/cm, with a total linear insertion loss of ~13 dB (5 dB/facet). Light was coupled in with tapered lensed fibers to SU8 polymer waveguides with inverse tapers.

**Phase-resolved characterization method.** For the nonlinear experiments, we used a mode-locked-laser (Alnair) fed into a pulse shaper (Finisar) generating near transform-limited 1.3 ps pulses at 1550 nm at a 30 MHz repetition rate. These pulses were then input into the FREG apparatus. The pulses were split into two branches by a fiber-coupler, with the majority of the energy coupled into the PhC-wg. The remaining fraction was sent to a reference branch with a variable delay, before being detected by a fast photodiode and transferred to the electronic domain. This electronic signal drove a Mach-Zender modulator that gated the optical pulse output from the PhC-wg. Using an optical spectrum analyser (OSA), we measured the spectra as a function of delay to generate a series of optical spectrograms. We de-convolved the spectrograms with a numerical algorithm (256 x 256 grid - retrieval errors G < 0.005), to retrieve the pulse intensity and the phase in both the temporal and spectral domain [25].

**Generalized Nonlinear Schrödinger Equation Model.** The parameters used in our GNLSE model for the slow-light dispersion engineered PhC-wg are: slow-down factor $S=n_g/n_0=8.6376$, effective linear absorption $\alpha_{l,eff}$ = 13.9 cm$^{-1}$; $\beta_2$=+1 ps$^2$/mm, a TOD parameter of $\beta_3$=0.02 ps$^3$/mm, and a FOD parameter of $\beta_4$=-2.2 ps$^4$/mm; effective nonlinear parameter $\gamma_{eff} = \frac{2\pi n_2}{\lambda_0 A_{eff}} S^2$ =4072 (W m)$^{-1}$, with $n_2$=6×10$^{-18}$ m$^2$/W and $A_{eff}$=0.44μm$^2$; effective TPA parameter $\alpha_{TPA,eff} = \frac{\beta_{TPA}}{A_{eff}} S^2 = 1674$ (W m)$^{-1}$, with $\beta_{TPA}$=10×10$^{-12}$ m/W; effective free-carrier dispersion parameter $n_{FC,eff}$=-6×10$^{-27}$ S m$^3$; effective free-carrier absorption parameter $\sigma_{eff}$= 1.45×10$^{-21}$ S m$^2$. The simulation results in Fig. 2 were obtained by using the measured input pulse as the input to the model. The simulation results in Fig. 4 were obtained using a perfect Gaussian, hyperbolic secant, and a super Gaussian (order 4) pulse of the same width as the experimental pulse, 1.3 ps. The linear loss in the nanowires that couple light into and out of the PhC-wg was negligible. Nonlinear absorption in the coupling nanowire (effective area, ~ 0.2 μm$^2$) was taken into account in the NLSE model.

**Acknowledgements**

This work was supported in part by the Center of Excellence CUDOS (CE110001018), Laureate Fellowship (FL120100029) schemes of the Australian Research Council (ARC). A. B.-R. was partially supported by the collaborative research project between The University of Sydney and Technion Israel Institute of Technology. T.F.K. was supported by EPSRC UK Silicon Photonics (Grant reference EP/F001428/1). C.H. was supported by the ARC Discovery Early Career Researcher award (DECRA - DE120102069).

# Supplementary Material

In this section we derive an analytic expression approximately describing the canonical form of a *fundamental pure-quartic soliton*. The canonical system, in the presence of just negative fourth-order dispersion (FOD), i.e. $\beta_4<0$, and self-phase-modulation (SPM) is described by the equation

$$i\frac{\partial A}{\partial z} = \frac{|\beta_4|}{24}\frac{\partial^4 A}{\partial t^4} - \gamma_{eff}|A|^2 A. \tag{1}$$

The experimental observations and numerical simulations previously presented in this paper indicate that a Gaussian is a good approximation to the form of the *fundamental pure-quartic soliton*. Next, we look for an approximate solution to Eq. (1) in two separate ways: using the variational principle and looking for a local approximation.

**Variational approximation to fundamental pure-quartic-solitons**

Assuming that a Gaussian is a good approximation, we can consider a related problem defined by the eigenvalue equation:

$$\mathcal{H}(t)\psi(t) = E\psi(t), \tag{2}$$

where the Hamiltonian operator $\mathcal{H}(t)$ is given by

$$\mathcal{H}(t) = \frac{|\beta_4|}{24}\frac{\partial^4}{\partial t^4} - \gamma_{eff}A_0^2 e^{-2\sigma t^2}, \tag{3}$$

with $A_0$ and $\sigma$ being real and positive numbers, $E$ being the eigenvalue, and $\psi(t)$ a given wave function. The exponential factor with $\sigma$ is the nonlinear refractive index profile of the soliton, but we do not use this information yet. Since $\mathcal{H}(t)$ is a Hermitian operator we can look for an approximate solution to Eq. (2) using the variational principle. We pick a trial wave function of the form

$$\psi(t) \propto e^{-\delta t^2}, \tag{4}$$

where $\delta$ is the variational parameter we will adjust. The expectation value of the energy term is then

$$E(\delta) = \frac{\frac{|\beta_4|}{24}\int \psi^*(t)\frac{\partial^4}{\partial t^4}\psi(t)dt - \gamma_{eff}A_0^2\int e^{-2\sigma t^2}|\psi(t)|^2 dt}{\int |\psi(t)|^2 dt}. \tag{5}$$

Doing two partial integrations, and taking into account that $\psi(t)$ is real, Eq. (5) can be simplified to

$$E(\delta) = \frac{\frac{|\beta_4|}{24}\int \left(\frac{\partial^2\psi(t)}{\partial t^2}\right)^2 \psi(t)dt - \gamma_{eff}A_0^2\int e^{-2(\sigma+\delta)t^2}dt}{\int e^{-2\delta t^2}dt}, \tag{6}$$

where we have used the fact that the result is independent of the normalization of $\psi(t)$, so we can take $\psi(t) = e^{-\delta t^2}$. Substituting $\frac{\partial^2\psi(t)}{\partial t^2}$ in Eq. 6, we get

$$E(\delta) = \frac{\frac{|\beta_4|}{24}\left(4\delta^2\int e^{-2\delta t^2}dt + 16\delta^4\int t^4 e^{-2\delta t^2}dt - 16\delta^3\int t^2 e^{-2\delta t^2}dt\right) - \gamma_{eff}A_0^2\int e^{-2(\sigma+\delta)t^2}dt}{\int e^{-2\delta t^2}dt}. \tag{7}$$

Evaluating the integrals we find

$$E(\delta) = \frac{|\beta_4|\delta^2}{8} - \gamma_{eff} A_0^2 \frac{\delta^{\frac{1}{2}}}{(\delta+\sigma)^{\frac{1}{2}}}. \tag{8}$$

According to the variational principle, for fixed $A_0$ and $\sigma$, i.e. for fixed Hamiltonian $\mathcal{H}(t)$, the best solution of the form we have adopted will be found by forcing $\frac{dE(\delta)}{d\delta} = 0$. This yields

$$\frac{dE(\delta)}{d\delta} = \frac{|\beta_4|\delta}{4} - \frac{1}{2}\gamma_{eff} A_0^2 \frac{1}{\delta^{\frac{1}{2}}(\delta+\sigma)^{\frac{1}{2}}} + \frac{1}{2}\gamma_{eff} A_0^2 \frac{\delta^{\frac{1}{2}}}{(\delta+\sigma)^{\frac{3}{2}}} = 0, \tag{9}$$

that determines the $\delta$ value that minimizes the energy of the trial wave function, i.e. the "best $\delta$". However, we are interested in knowing, under which circumstances the "best $\delta$" equals $\sigma$. By setting $\delta = \sigma$ in Eq. (9) we get to

$$\sigma^2 = \frac{\gamma_{eff}}{\sqrt{2}|\beta_4|} A_0^2, \tag{10}$$

which determines the conditions under which the "best $\delta$" equals $\sigma$. In this special case, our estimate for the energy (8) is

$$E(\delta) = -\frac{7}{8} \frac{\gamma_{eff} A_0^2}{\sqrt{2}}. \tag{11}$$

Using (3), (10) and (11) we can write a special case of Eq. (2) as

$$\left(\frac{|\beta_4|}{24}\frac{\partial^4}{\partial t^4} - \gamma_{eff} A_0^2 e^{-2\left(A_0\sqrt{\frac{\gamma_{eff}}{\sqrt{2}|\beta_4|}}\right)t^2}\right)\left(A_0 e^{-\left(A_0\sqrt{\frac{\gamma_{eff}}{\sqrt{2}|\beta_4|}}\right)t^2}\right) \approx \left(-\frac{7}{8}\frac{\gamma_{eff} A_0^2}{\sqrt{2}}\right)\left(A_0 e^{-\left(A_0\sqrt{\frac{\gamma_{eff}}{\sqrt{2}|\beta_4|}}\right)t^2}\right) \tag{12}$$

Next we return to Eq. (1) and look for an approximate solution of the form:

$$A(z,t) = A_0 e^{i\Gamma z} e^{-\left(A_0\sqrt{\frac{\gamma}{\sqrt{2}|\beta_4|}}\right)t^2}. \tag{13}$$

Substituting Eq. (13) in Eq. (1) we find

$$-\Gamma\left(A_0 e^{i\Gamma z} e^{-\left(A_0\sqrt{\frac{\gamma_{eff}}{\sqrt{2}|\beta_4|}}\right)t^2}\right) = \left(\frac{|\beta_4|}{24}\frac{\partial^4}{\partial t^4} - \gamma_{eff} A_0^2 e^{-2\left(A_0\sqrt{\frac{\gamma_{eff}}{\sqrt{2}|\beta_4|}}\right)t^2}\right)\left(A_0 e^{i\Gamma z} e^{-\left(A_0\sqrt{\frac{\gamma_{eff}}{\sqrt{2}|\beta_4|}}\right)t^2}\right). \tag{14}$$

Using Eq. (12) and Eq. (14) we find

$$-\Gamma\left(A_0 e^{i\Gamma z} e^{-\left(A_0\sqrt{\frac{\gamma_{eff}}{\sqrt{2}|\beta_4|}}\right)t^2}\right) \approx \left(-\frac{7}{8}\frac{\gamma_{eff} A_0^2}{\sqrt{2}}\right)\left(A_0 e^{i\Gamma z} e^{-\left(A_0\sqrt{\frac{\gamma_{eff}}{\sqrt{2}|\beta_4|}}\right)t^2}\right), \tag{15}$$

which yields

$$\Gamma = \frac{7}{8}\frac{\gamma_{eff} A_0^2}{\sqrt{2}}. \tag{16}$$

We finally get to a class of approximate solutions to Eq. (1), characterized by their maximum input amplitude $A_0$ of the form

$$A(z,t) = A_0 e^{i\left(\frac{7\gamma_{eff}A_0^2}{8\sqrt{2}}\right)z} e^{-\left(A_0\sqrt{\frac{\gamma_{eff}}{\sqrt{2}|\beta_4|}}\right)t^2}. \tag{17}$$

**Local approximation to fundamental pure-quartic-solitons**

Here, we find an approximate solution to Eq. (1) using a local approximation. Since we know a Gaussian is a good approximation we take

$$A(z,t) = A_0 e^{i\Gamma z} e^{-\frac{t^2}{\tau^2}}, \tag{18}$$

which is equivalent to Eq. (3) in the main text. Substituting Eq. (18) in Eq. (1) and Taylor expanding $e^{-\frac{t^2}{\tau^2}}$ we get to

$$-\Gamma A_0 \left(1 - \frac{t^2}{\tau^2} + \frac{t^4}{2\tau^4}\right) = \frac{|\beta_4|}{24}\frac{4}{\tau^4}\left(3 - 12\frac{t^2}{\tau} + 4\frac{t^4}{\tau}\right)\left(1 - \frac{t^2}{\tau^2} + \frac{t^4}{2\tau^4}\right) A_0 - \gamma_{eff}\left(1 - 3\frac{t^2}{\tau^2} + \frac{9}{2}\frac{t^4}{\tau^4}\right) A_0^3. \tag{19}$$

Solving Eq. (19) to order $t^0$ and $t^2$, respectively, yields

$$\Gamma = \gamma_{eff} A_0^2 - \frac{1}{2}\frac{|\beta_4|}{\tau^4} \tag{20}$$

and

$$\Gamma = 3\gamma_{eff} A_0^2 - \frac{5}{2}\frac{|\beta_4|}{\tau^4}. \tag{21}$$

The intersection between Eq. (20) and Eq. (21) determines the parameters $\tau$ and $\Gamma$ for this approximation

$$\tau = \sqrt[4]{\frac{|\beta_4|}{\gamma A_0^2}}, \tag{22}$$

$$\Gamma = 3\gamma_{eff} A_0^2 - \frac{5}{2}\frac{|\beta_4|}{\tau^4}. \tag{23}$$

Therefore, using a local method, we get to a class of approximate solutions to Eq. (1), characterized by their maximum input amplitude $A_0$ of the form

$$A(z,t) = A_0 e^{i(\frac{1}{2}\gamma_{eff}A_0^2)z} e^{-\left(A_0\sqrt{\frac{\gamma_{eff}}{|\beta_4|}}\right)t^2}. \tag{24}$$

The fact that the variational and local approximations give very similar results, as illustrated by Eq. (17) and Eq. (24), together with the good matching they provide to the numerical solution of Eq. (1), reinforces our confidence in them.